\documentclass[aps,prl,groupedaddress,twocolumn,superscriptaddress,amsfonts,amssymb,amsmath,showpacs,floatfix,preprintnumbers]{revtex4-2}

\usepackage{dcolumn}
\usepackage{bm}
\usepackage{multirow}
\usepackage{verbatim}
\usepackage{graphicx}
\usepackage[normalem]{ulem}
\usepackage{color}
\usepackage{placeins}
\usepackage{xcolor}
\usepackage{slashed}
\usepackage{physics}
\usepackage{hyperref}

\newcommand{\msbar}{\overline{\rm MS}}

\begin{document}

\preprint{JLAB-THY-26-4922}

\title{Femtoscale imaging of the proton with
Ioffe-time distributions}



\author{ Robert G. Edwards}
\email[e-mail: ]{edwards@jlab.org}
\affiliation{Thomas Jefferson National Accelerator Facility, Newport News, VA 23606, USA}

\author{ Joe Karpie}
\email[e-mail: ]{jkarpie@jlab.org}
\affiliation{Thomas Jefferson National Accelerator Facility, Newport News, VA 23606, USA}

\author{ Christopher Monahan}
\email[e-mail: ]{cmonahan2024@coloradocollege.edu}
\affiliation{Department of Physics, Colorado College, Colorado Springs, CO 80903, USA}

\author{ Kostas Orginos}
\email[e-mail: ]{kostas@wm.edu}
\affiliation{Physics Department, William \& Mary, Williamsburg, VA 23187, USA}

\author{ Anatoly Radyushkin}
\email[e-mail: ]{radyush@jlab.org}
\affiliation{Department of Physics, Old Dominion University, Norfolk, Virginia, USA}
\affiliation{Thomas Jefferson National Accelerator Facility, Newport News, VA 23606, USA}

\author{ David Richards}
\email[e-mail: ]{dgr@jlab.org}
\affiliation{Thomas Jefferson National Accelerator Facility, Newport News, VA 23606, USA}

\author{ Eloy Romero}
\email[e-mail: ]{eloy.romero-alcalde@cpt.univ-mrs.fr}
\affiliation{Aix Marseille Univ, Universit\'e de Toulon, CNRS, CPT, Marseille, France.}

\author{ Savvas Zafeiropoulos}
\email[e-mail: ]{savvas.zafeiropoulos@cpt.univ-mrs.fr}
\affiliation{Aix Marseille Univ, Universit\'e de Toulon, CNRS, CPT, Marseille, France.}

\begin{abstract}
Mapping how strongly interacting constituents are distributed within protons is a key goal of nuclear physics
 and a  major  direction of  the future Electron Ion Collider program. 
 We propose a novel space-time description of hadron structure in terms of impact-parameter Ioffe-time distributions,
 relating spatial density in the  plane transverse to the proton momentum and the time between the probe's absorption and the product's emission
 in the longitudinal direction. Using lattice Quantum Chromodynamics, we perform the  first calculation of the Ioffe-time-dependent mean squared proton radii and compare our results  with  estimates in the Goloskokov-Kroll model. 
  We derive a relationship between the experimentally measurable Compton form factor and the generalized  Ioffe-time distribution, which allows us to perform the first  extraction of the  Compton form factor from lattice calculations.
\end{abstract}

\maketitle

\section{Introduction}

Understanding how quarks and gluons are distributed inside hadrons is one of the central goals of modern nuclear physics and a primary scientific objective of the Jefferson Lab 12 GeV program and the future 
Electron Ion Collider (EIC). While Quantum Chromodynamics (QCD)  provides the fundamental theory of the strong interaction, the nonperturbative structure of hadrons cannot in general be computed analytically and must instead be determined through experiment and first-principles numerical calculations.

Traditionally, the partonic structure of hadrons
in inclusive reactions, 
such as Deep Inelastic Scattering (DIS) and Drell-Yan (DY) process, 
  is described  by parton distribution functions (PDFs) $f(x)$ 
  specifying the fraction of the total energy and momentum of the hadron carried by a specific parton.

   Generalized parton distributions (GPDs)  $H(x,\xi;t)$, etc.,   extend this partonic description to
   hard exclusive processes,  such as 
    deeply virtual Compton scattering (DVCS) ~\cite{Muller:1994ses,Ji:1996nm,Radyushkin:1996nd}, deeply virtual meson production (DVMP)~\cite{Radyushkin:1996ru,Collins:1996fb}, and  single diffractive hard exclusive processes (SDHEP)~\cite{Qiu:2022pla}, which are 
targets for current and future experiments at 
CEBAF~\cite{Dudek:2012vr,Accardi:2023chb}, SPS~\cite{Gautheron:1265628,Adams:2676885}, J-PARC~\cite{Sawada2016} and EIC~\cite{Accardi:2012qut}.

The extra two variables $\xi, t$ of GPDs specify the momentum $q$ transferred from the initial to the final hadron 
involved in the hard exclusive reaction through its longitudinal $\xi$ and total $t$  magnitude. The Fourier-conjugate ${\bf b}_T$ of the transverse component of $q$, ${\bf q}_T$, is called the {\it impact parameter}.  

The impact parameter distribution $h(x, b_T)$ (IPD)~\cite{Burkardt:2002ks}
 employs  $b_T$ to describe
the transverse structure of the hadron in the plane orthogonal to the direction of the hadron momentum $p$  while using the momentum fraction $x$ to describe the structure in the longitudinal direction. 
Long ago Braun, G\'ornicki, and 
Mankiewicz~\cite{Braun:1994jq} proposed switching from momentum fraction $x$ and PDFs to a description in terms of the parameter $\nu$, the Fourier-conjugate to $x$. They called this parameter the ``Ioffe-time", 
and introduced the Ioffe-time distributions ${\cal I}(\nu)$ (ITD), 
shown later~\cite{Braun:2007wv,Radyushkin:2017cyf} to be the  natural object for lattice calculations. 

In this work, we introduce 
the impact-parameter Ioffe-time distributions (IP-ITD). We argue that  IP-ITDs provide natural and intuitive representations of the three-dimensional structure of hadrons by using a completely space-time description. 
For instance, in DVCS a parton absorbs the virtual photon probe for some amount of time before a real photon is emitted. 
Note that unlike the original Ioffe-time scale $\tau \sim 1/x_B$~\cite{Ioffe:1969kf}, the Ioffe-time 
$\nu=p\cdot z$ of  Ref.~\cite{Braun:1994jq}, 
is a precisely-defined interaction time, rather than an order of magnitude estimate. 

Thus, in nuclear femtography, one images the proton's transverse spatial distribution in terms of the parton's Ioffe-time, i.e.~the interaction time.
In this description, {\it flash photography}, via DVCS, describes the proton through an intuitive spacetime view of how quarks behave in each image.

The dimensionless 
parameters $\tau$ and $\nu$
measure time in units of the hadron's mass, for protons $m_{p}^{-1}\sim 0.7$ yocto-seconds (1~ys = $10^{-24}$~s) and pions $m_{\pi}^{-1}\sim 4.7$ ys~\cite{ParticleDataGroup:2024cfk}. 
For comparison the timescales
of the average lifetimes, $t= 1/\Gamma$, of the $Z$ boson $t_Z= 0.26$ ys, the $W$ boson $t_W=0.31$ ys, and the top quark $t_t=0.46$ ys 
etc.~\cite{ParticleDataGroup:2024cfk}.

In this paper, we define the impact-parameter Ioffe-time distributions and  provide the first numerical estimate of the Ioffe-time-dependent transverse radii of the proton from lattice QCD and the popular phenomenological Goloskokov-Kroll (GK) model~\cite{Goloskokov:2007nt}. We also show that the Compton form factors (CFFs),
the functions through which GPDs reveal themselves in experiments, may be written 
 in terms of Ioffe-time distributions and, for the first time, we extract CFF from lattice QCD. 


\section{Spatial Parton distributions\label{sec:parton_defs}}

We begin   with  basic information about 
ITDs.  In particular, we present  
 definitions of momentum fraction distributions  in terms of ITDs.

The starting point of spatial imaging of the proton is 
the off-forward nucleon matrix element of quark fields separated by a light-cone distance $z^\mu$
\begin{align}
    &\langle p_f, \lambda_f  | \bar{\psi}(-z/2) \gamma^+ W(-z/2,z/2) \psi(z/2) | p_i, \lambda_i\rangle \nonumber\\ &
    =\frac{1}{P^+} \bar{u}(p_f,\lambda_f) \left[ \gamma^+ \mathcal{H}(\nu,\bar\nu,t) + \frac{i\sigma^{+\nu}q_\nu}{2m} \mathcal{E}(\nu,\bar\nu,t)\right.  \nonumber\\ & \hspace{3cm}  \left. - \frac{q^+}{2m} \mathcal{D}(\bar\nu,t)\right] u(p_i, \lambda_i)\label{eq:gpd_def}\,,
\end{align}
where    $q^\mu \equiv p_f^\mu-p_i^\mu$ is the momentum transfer, \mbox{$P^\mu \equiv \frac{1}{2}(p_i^\mu+p_f^\mu)$} 
is the average momentum and $u(p,\lambda)$ are the normalized nucleon spinors.

The generalized Ioffe-time distributions (GITD) $\mathcal{H}$, $\mathcal{E}$  and $\mathcal{D}$  defined by the Lorentz decomposition of 
the  matrix element 
are functions of the Lorentz invariants  $\nu \equiv P\cdot z$, $\bar{\nu} \equiv -q\cdot z/2$  and  \mbox{$t \equiv q^2$}.
 In renormalizable theories, GITDs also depend on a factorization scale $\mu^2$, suppressed for brevity.
 
As functions of two independent  variables 
$\nu$ and $\bar \nu$, GITDs $ \mathcal{H}, \mathcal{E}$
may be written in the  form of a  Fourier transform  of  {\it double distributions} (DDs)~\cite{Muller:1994ses,Radyushkin:1996nd}, 
\begin{equation}
    \begin{pmatrix}
        \mathcal{H}(\nu,\bar\nu,t)\\
        \mathcal{E}(\nu,\bar\nu,t)
    \end{pmatrix} =  \int_\Omega \dd \beta\dd \alpha\,\,e^{i\beta \nu + i\alpha \bar \nu}  
    \begin{pmatrix}
        h(\beta,\alpha,t)\\
        e(\beta,\alpha,t)
    \end{pmatrix} \,  ,
      \label{DDdef}
\end{equation} 
where the support region $\Omega$  for  DD  is   a rhombus
 specified by
$|\alpha|+|\beta| \leq 1$ \cite{Radyushkin:1998bz}. 

 The skewness $\xi\equiv \bar\nu/\nu$
 characterizes the  size of the  longitudinal momentum transfer.  Alternatively, choosing  $\nu$ and $\xi $
as independent  variables and taking the Fourier transform in $\nu$ gives the GPDs. 
Combining the definition 
\begin{equation}
    \begin{pmatrix}
        \mathcal{H}(\nu,\xi\nu,t)\\
        \mathcal{E}(\nu,\xi\nu,t)
    \end{pmatrix} = \int_{-1}^1 dx\, e^{i\nu x} 
    \begin{pmatrix}
        H(x,\xi,t)\\
        E(x,\xi,t)
    \end{pmatrix} 
      \label{HEdef}
\end{equation}    
with eq.~(\ref{DDdef}) 
produces a direct relation between GPDs and DDs~\cite{Muller:1994ses,Radyushkin:1998bz},
\begin{align}
    \begin{pmatrix}
        H(x,\xi,t)\\
        E(x,\xi,t)
    \end{pmatrix} =&
    \int_{-1}^1 d\beta \int_{-1+|\beta|}^{1-|\beta|} d\alpha \,
   \nonumber \\ & \times  \delta (x -(\beta +\xi \alpha))
    \begin{pmatrix}
       h(\beta, \alpha,t)\\
       e(\beta, \alpha,t)
    \end{pmatrix} 
\,.
      \label{HEDD}
\end{align}

GPDs have two distinct regions: the Dokshitzer-Gribov-Lipatov-Altarelli-Parisi (DGLAP)~\cite{Gribov:1972ri,Altarelli:1977zs,Dokshitzer:1977sg} region $|x|>|\xi|$, where GPDs look like distorted PDFs, and the Efremov-Radyushkin-Brodsky-Lepage (ERBL)~\cite{Efremov:1979qk,Lepage:1980fj} region $|x|<|\xi|$, where GPDs look like distorted distribution amplitudes. At the border points 
 $x=\pm \xi$, GPDs are non-analytic functions of $x$.  
In contrast, GITDs are given by simple Fourier integrals of DDs,
and are smooth functions of $\nu$ and $\bar \nu$.

The DD representation, 
eq.(\ref{HEDD}),
automatically imposes {\it polynomiality}~\cite{Ji:1998pc} on GPDs: the $x^n$ moments of $H$ and $E$ are  $n^{\rm th} $   order  polynomials in $\xi$.
In lattice QCD calculations of Ref.~\cite{Dutrieux:2026grg},
 DDs have been   used to extract GPDs   that automatically satisfy the polynomiality condition. 

Under the GPD convention \mbox{$q^+= - 2\xi P^+$}  one may use the Gordon decomposition to combine the \mbox{$\mathcal D$-term}~\cite{Polyakov:1999gs} with the  $\mathcal H$  and $\mathcal E$ contributions. This changes 
\mbox{$H\to H+\xi D$}  and $E\to E -\xi D$ for GPDs.
As a result,  odd-$n$ $x^n$ moments of ``full'' 
$H(x,\xi)$  and $E(x,\xi) $ GPDs are the \mbox{$(n+1)^{\rm th} $} order polynomials of $\xi$,
with 
the $\xi^{n+1}$  term coming solely  from $D$.
Our GPD lattice  extractions are based on the 
DD-decomposition
(\ref{DDdef}), and \mbox{$\mathcal D$-term} is obtained separately from $\mathcal H$ and $\mathcal E$. For simplicity, in the remainder of the paper we focus on $H(x,\xi,t)$, but  analogous expressions and relationships can be derived for other GPDs.

The $x^n$ moments of the GPDs are directly related to the coefficients of the Taylor expansion, in $\nu$, of the GITD~\cite{Braun:2007wv,Karpie:2018zaz,Pang:2024kza}. For example, the gravitational form factors (GFFs), which encode the mechanical and spatial properties of the nucleon, are given by the first derivative. For the ``DD part'' of $H(x,\xi,t)$ we have 
\begin{equation}
         \int_{-1}^1 dx \, x^n \, H(x,\xi,t) = (-i)^n\frac{\partial^n}{\partial\nu^n} 
         \mathcal{H}(\nu,\xi\nu,t)\Big|_{\nu=0}\,.\label{eq:gffs}
\end{equation}
By polynomiality for $n=0$ or 1, the value of $\mathcal{H}$ and its first derivative at $\nu=0$ do not depend on $\xi$. 
This means that the lattice data for
 ${\rm Im} \ \mathcal{H}(\nu,\xi\nu,t)$ at various $\xi$ should have the same slope at $\nu=0$. This provides a test on the data and a constraint on the inverse problem.

The transverse spatial distribution of quarks can be described by the impact parameter distributions (IPD)~\cite{Burkardt:2002ks}
\begin{equation}
        h(x,b_T) = \int \frac{d^2 q_T}{(2\pi)^2} e^{-i q_T\cdot b_T} 
        H(x,0,-q_T^2)\,.
\end{equation}
The transverse size  can be most simply captured through 
the momentum-fraction-dependent radii
\begin{equation}
    r_H^2(x) = \frac{\int d^2b_T \,b_T^2 h(x,b_T)}{\int d^2b_T  h(x,b_T)}.
    \label{rx}
\end{equation}

We now introduce the impact-parameter Ioffe-time distributions (IP-ITD).
They  describe the $b_T$ distribution that is seen by a probe that interacts with the proton 
for a time $\nu/m_p$ and are defined through the Fourier transform of  IPDs 
\begin{align}
    h(\nu,b_T)  =& \int \frac{d^2 q_T}{(2\pi)^2} e^{-i q_T\cdot b_T} \mathcal{H}(\nu,0,-q^2_T) \nonumber\\
    =& \int dx\, e^{i\nu x} h(x,b_T)\,.
\end{align}

Operators with light-like separations cannot be accessed in Euclidean QCD. Instead, 
operators containing space-like-separated fields 
are used in the basic eq. (\ref{eq:gpd_def}). 
As a result, one deals with pseudo-GITDs
$\mathcal{H}(\nu,\bar\nu,t;z^2)$, etc., which  have an 
extra dependence on space-like interval $z^2$. 

The strongest part of the $z^2$-dependence is 
related to Wilson line renormalization and
nonperturbative \mbox{large-$z^2$} ``higher twist'' effects.
Its bulk part  is   eliminated  using   
{\it reduced}  pseudo-GITDs~\cite{Dutrieux:2026grg}, obtained by dividing 
pseudo-GITDs by their $\nu=0$ counterparts.
The remaining mild \mbox{$z^2$-dependence} at large $z^2$ is fitted. 

At  small $z^2$ one also deals with $z^2$-dependence 
 caused by QCD  evolution. It   is taken care of by
perturbative matching ~\cite{Dutrieux:2026grg}. 
For example,  reduced pseudo-GITD is related to  light-cone GITD via~\cite{Radyushkin:2019owq}
\begin{equation}
        \mathcal{H}(\nu,\bar{\nu}, t, z^2) = \int_0^1 d\alpha\, C(\alpha, \bar\nu, z^2\mu^2) 
        \mathcal{H}(\alpha\nu,\alpha\bar{\nu}, t, \mu^2) \,,
\end{equation} 
where $C(\alpha, \bar\nu, z^2\mu^2) $ is calculable in perturbation theory; an explicit expression 
may be found in~\cite{Dutrieux:2026grg}.  
The pseudo-GPD and GPD are related by a similar convolution, which enables the extraction of GITDs and GPDs in the $\msbar$ scheme at $\mu= 2$ GeV from lattice QCD calculations of the pseudo-distributions.

Lorentz-invariant definitions of Ioffe-time spatial distributions provide a significant advantage: one can interpret them in a frame-independent manner. Lattice calculations can be carried out in any frame and one still 
may interpret the results in the rest-frame of the proton, where the Ioffe-time measures a literal time separation rather than a separation along the light cone.

\section{Transverse Radius of the Proton\label{sec:radius}}

In analogy to  (\ref{rx}), we introduce the Ioffe-time-dependent transverse radius
\begin{equation}
        r_H^2(\nu) = \frac{\int d^2b_T \,b_T^2 \,h(\nu,b_T)}{\int d^2b_T  \,h(\nu,b_T)} = \frac{4 \frac{\partial}{\partial t}\mathcal{H}(\nu,0,t)}{\mathcal{H}(\nu,0,t)}\bigg|_{t=0}
        \label{rb}
\end{equation}
that characterizes the transverse spatial structure of the hadron in terms of the IP-ITD. We determine this radius using data from lattice QCD and compare the results to the GK model.

In Ref.~\cite{Dutrieux:2026grg}, the matrix elements defining the nucleon's iso-vector quark pseudo-GITD were calculated using lattice QCD on an ensemble of $N_f=2+1$ clover-improved Wilson fermions from the JLab/W\&M/LANL/MIT/Marseille collaboration~\cite{Yoo:2026uul}, with lattice spacing $a=0.094$ fm~\footnote{Note this lattice spacing differs from Ref.~\cite{Yoo:2026uul}. Since a continuum limit is not being taken, we use an older estimate for consistency with previous work on this ensemble by the HadStruc Collaboration.} and pion mass $m_\pi = 358$ MeV. At this quark mass, the proton mass is $m_p\sim$1.15 GeV~\cite{Khan:2020ahz} ($m_p^{-1}=0.57$ ys). That analysis did not include $\mathcal{H}(\nu,0,t=0)$ data from Ref.~\cite{Egerer:2021ymv}. The following shows the results with and without its inclusion.

Extracting the GPD from the pseudo-GITD is an ill-posed inverse problem, which was resolved in Ref.~\cite{Dutrieux:2026grg} via Gaussian Process Regression (GPR). GPR provides a robust framework for uncertainty quantification used in some form to determine $x$-dependence from Ioffe-time data in Refs~\cite{Karpie:2019eiq, Alexandrou:2020tqq, Candido:2024hjt, Dutrieux:2024rem, Dutrieux:2025jed, Xiong:2025obq, Medrano:2025cmg, Ling:2025olz}. 

In Ref.~\cite{Dutrieux:2026grg}, the $(\beta,\alpha,t)$ dependence of the pseudo-DD was modeled using a three-dimensional GPR with a prior distribution with vanishing mean and covariance that embedded expected behaviors at large and small $\beta$ and along the boundary of the domain. Parameterizing  GITDs in terms of DDs has several advantages: the relationship is very general, based solely on Lorentz invariance, and enforces GPD polynomiality intrinsically correlating $\nu$ and $\bar{\nu}$. The supplementary materials use the Kullback-Leibler divergence~\cite{Kullback:1951zyt} to show the sensitivity in $(\alpha,\beta)$ and discuss impact on extrapolations in $(\nu,\bar{\nu})$.

\begin{figure}
    \includegraphics[width=0.95\linewidth]{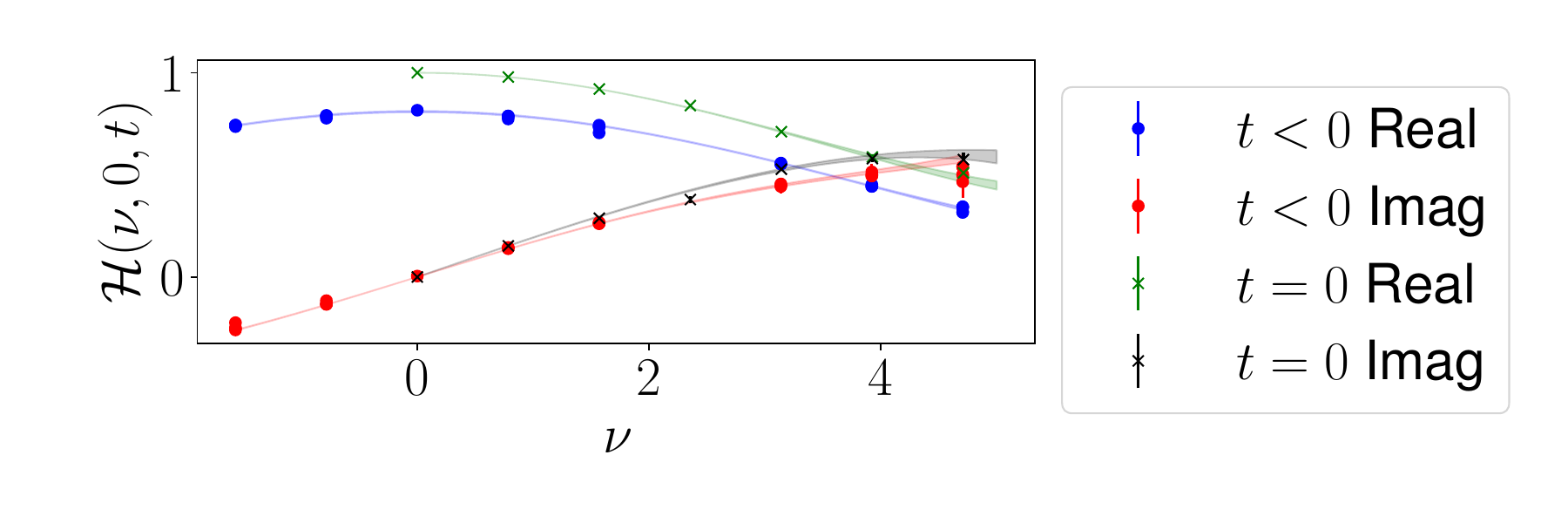}
    \caption{Fits of the lattice QCD datasets at $\bar\nu=0$ with $t=0$ and $t\in[-0.17,-0.16]$ GeV${}^2$. For the forward case, the real and imaginary analyses have reduced $\chi^2$ of 0.95 and 2.0, while the off-forward case has 1.18 and 2.0. The forward imaginary datum at $p_z=3\frac{2\pi}L$ suffers from excited state contamination causing underestimation and inflating $\chi^2$. Instead, the off-forward imaginary dataset has statistically incompatible points, inflating $\chi^2$, at both ends of the $\nu$ range likely due to other lattice systematics. \label{fig:poly_fits}}
\end{figure}

In Ioffe-time space, polynomiality is manifest as constraints on the $\xi$ dependence of the Taylor expansion of the GITD with respect to $\nu$~\cite{HadStruc:2024rix,Gao:2025inf}. Fig.~\ref{fig:poly_fits} shows data for the forward matrix element and an off-forward bin, at $\bar\nu=0$, fit  to even/odd polynomials in $\nu$ with three parameters.

\begin{figure}
    \centering
    \includegraphics[width=0.95\linewidth]{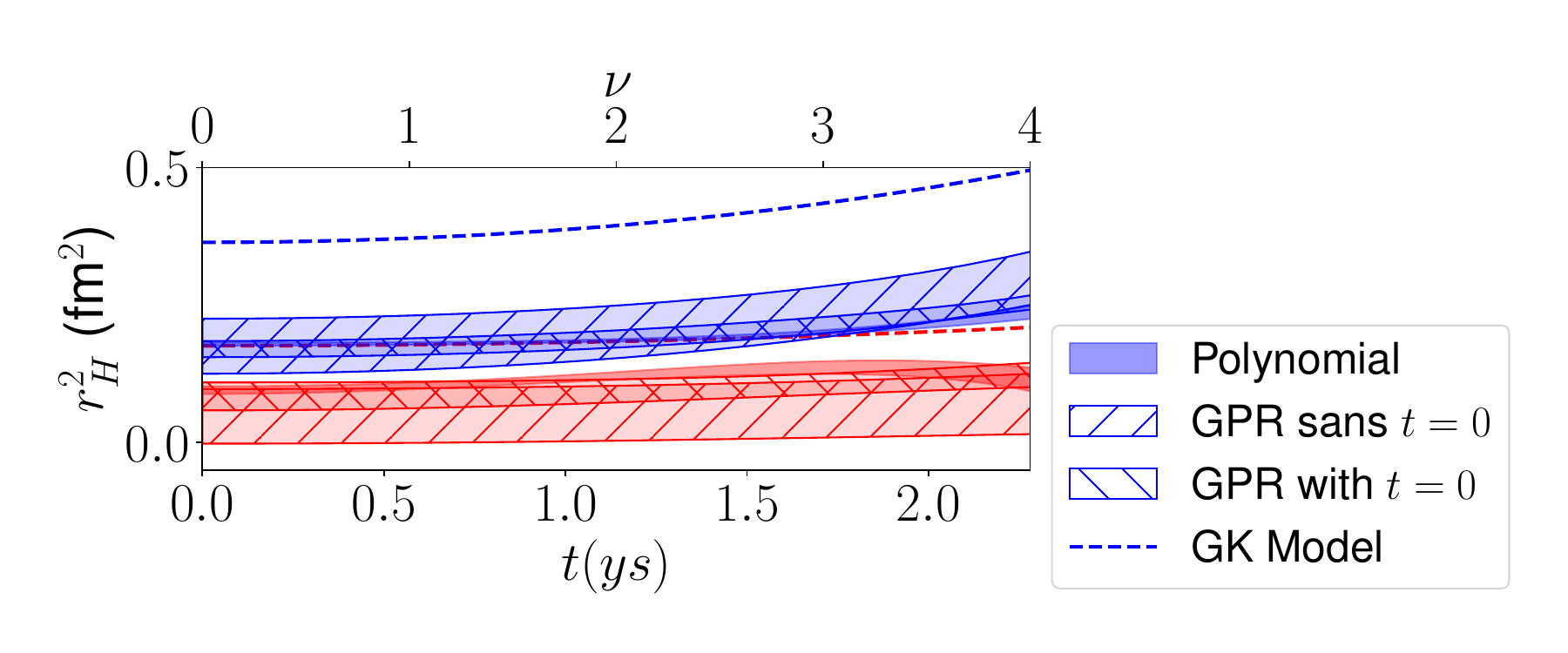}
    \caption{The Ioffe-time dependent transverse $r_H$ radius of the proton's iso-vector quarks from lattice QCD (bands) and the GK model (dashed lines). The $CP$ even (blue) and odd components (red) are in agreement. The upper scale measures the invariant $\nu$ while the lower measures time in the lattice nucleon's units.}
    \label{fig:itd_radius_H}
\end{figure}

GITD results are used to estimate the $t=0$ derivative, i.e. transverse spatial radii from lattice QCD, shown in Fig.~\ref{fig:itd_radius_H}. For the polynomial fits, we also estimate the radii via the difference of the fits. For GPR, we use the finite difference of the matched GITD at $t=0$ and a small $t=-0.2$ GeV${}^2$ to estimate the derivative with respect to $t$.

For the $CP$-even combination of iso-vector quarks, the radius is larger than for the $CP$-odd one. Shorter interaction times lead to smaller transverse distributions while longer interaction times probe wider $b_T$, in analogy to the transverse distribution is narrowing as $x$ increases.

Alongside, the GK model is shown. It  gives larger transverse radii while sharing the same trend. The heavy quarks in the lattice QCD calculation could potentially lead to smaller pion clouds and future studies will focus on this chirally sensitive behavior. Modeling the $t$-dependence through a dipole fit or $z$-expansion improves future analyses since finite differences are expected to underestimate the radii.

\section{Compton Form Factors and Generalized Ioffe-time distributions}
\label{sec:cff}

The simplest  process that requires GPD  description is DVCS.
The GPDs appear  in an  integrated form, through 
CFF given by  
\begin{align}
    \mathfrak{H}(\xi,t) = \int_{-1}^1 \dd x \left[\frac{1}{x- \xi+ i\epsilon }+\frac{1}{x+\xi-i\epsilon }\right] H(x,\xi,t)\, .
    \label{CFF}
\end{align}
Clearly, for a fixed $t$ one cannot reconstruct in a model-independent way a function of two variables $x,\xi$ 
from a function of just one variable $\xi$.

In this respect, the lattice calculations
are in much better position, since they allow one to extract 
 GPD $H(x,\xi,t)$ directly,  using  
 quasi-GPD~\cite{Liu:2019urm, Chen:2019lcm, Alexandrou:2020zbe}, the OPE-without-OPE~\cite{CSSMQCDSFUKQCD:2021lkf,Hannaford-Gunn:2024aix} or, as in this work, pseudo-GPD~\cite{Radyushkin:2019owq, Dutrieux:2026grg} approaches. 
Still, to compare the results with the DVCS data
one should obtain CFFs using eq. (\ref{CFF}). 
However, the CFF integral has  singularities for \mbox{$x=\pm \xi$} that can  adversely affect the accuracy of the
CFF calculation. In the quasi-PDF approach, one 
should also ensure that extracted  GPDs  $H(x,\xi,t)$
corresponding to different $\xi$ obey the 
polynomiality condition.

Fortunately, in this approach,  the CFF $ \mathfrak{H}(\xi,t)$ can be obtained directly from  GITD $\mathcal{H}(\nu,\xi\nu,t)$,  by-passing the extraction of GPD $H(x,\xi,t)$ and replacing integration over $x$ with $\nu$.

Proceeding with the proof, notice that 
 only the odd part $ H^{(+)} (x,\xi)\equiv  H(x,\xi)- H(-x,\xi)$  contributes to eq.~\eqref{CFF}.
By time reversal invariance, the  double ITD  $\mathcal{H}(\nu,\bar  \nu,t) $ of eq.~\eqref{eq:gpd_def}
is even in $\bar  \nu$, but it has both even 
$ \mathcal{H}^{(-)} (\nu,\bar  \nu,t)$  and odd  $\mathcal{H}^{(+)} (\nu,\bar  \nu,t) $  components with respect to its first argument $\nu$, so that $ \mathcal{H}(\nu,\bar  \nu,t) = \mathcal{H}^{(-)} (\nu,\bar  \nu,t)+i \mathcal{H}^{(+)} (\nu,\bar  \nu,t)$.
  Using the definition  of  GPDs, eq.~\eqref{HEdef},
 we  write 
\begin{align}
   \mathcal{H}^{(+)} (\nu,\xi\nu,t)  =  \int_{0}^1  \dd x \,  \sin (x\nu) \,    {H}^{(+)} (x,\xi,t)  \ .
    \label{CFF4a}
\end{align}
Incorporating 
\begin{align}
 \frac{1}{x- \xi+ i\epsilon }+\frac{1}{x+\xi-i\epsilon } = 2 \int_{0}^\infty \dd\nu\,  \sin (x \nu) e^{-i \xi  \nu}  \, e^{-\epsilon \nu}  \ ,
\end{align}
we  derive  
\begin{align}
    \mathfrak{H}(\xi,t) =& 2\int_{0}^1 \dd x \,   \int_{0}^\infty \dd\nu\,  \sin (x \nu) e^{-i \xi  \nu}  \, e^{-\epsilon \nu} H^{(+)}(x,\xi,t)\,.
    \label{CFFsin}
\end{align}
 The $x$-integral here gives the GITD~\eqref{CFF4a},  
and we have 
\begin{align}
    \mathfrak{H}(\xi,t) =& 2    \int_{0}^\infty \dd\nu\,   e^{-i \xi  \nu}  \, e^{-\epsilon \nu}   \mathcal{H}^{(+)} (\nu,\xi\nu,t) \,.
    \label{Kerrep2}
\end{align}
This relation gives a  CFF $\mathfrak{H}(\xi,t)$  as a simple  
integral  of a GITD $ \mathcal{H}^{(+)}(\nu,\xi\nu,t)$. 
Note that  ``$\xi$'' is present here 
 not only  in  the 
$e^{-i \xi\nu}  $  factor, 
but also among the arguments of $\mathcal{H}^{(+)}(\nu,\xi\nu,t)$. 
Hence,  eq.~\eqref{Kerrep2}  cannot  be  Fourier-inverted
 to express $   \mathcal{H}^{(+)}(\nu,\xi\nu,t)$ 
in terms of $ \mathfrak{H}(\xi,t)$.

Since $ \mathcal{H}^{(+)} (\nu,\xi\nu,t)$  vanishes when $\nu \to \infty$, we may omit the  $e^{-\epsilon \nu} $ factor, obtaining  
\begin{align}
{\rm Im}  \,   \mathfrak{H}(\xi,t) =- {2}   \int_{0}^\infty \dd\nu\,  \sin(\xi \nu )  \, 
\mathcal{H}^{(+)}(\nu,\xi\nu,t) \ 
    \label{CFFpIm}
\end{align}
for the  imaginary part.
By the inverse of eq. (\ref{CFF4a}),  it  is equal to the well-known  [and evident from eq.(\ref{CFF})]
$ -i \pi H^{(+)}(\xi, \xi, t)$ contribution to the CFF. It  involves the GPD $H^{(+)}(x, \xi, t)$ taken at the  border point $x=\xi$
 between the DGLAP and ERBL regions. The relation 
\begin{align}
&
 {\rm Re}  \,   \mathfrak{H}(\xi,t) = 2 \int_{0}^\infty  \dd \nu \,  \cos (\xi\nu) \,  \mathcal{H}^{(+)} (\nu,\xi\nu,t) \ 
    \label{CFFpRe}
\end{align}
gives the  real part of the CFF. 
It substitutes a  cumbersome  principal-value integral of $H^{(+)}(x,\xi;t)/(x-\xi)$. 

The crucial advantage of Eqs. ~\eqref{CFFpIm} and \eqref{CFFpRe} is that they  can be used to directly 
calculate CFF $\mathfrak{H}(\xi,t)$   from the lattice results for GITD $\mathcal{H}^{(+)}(\nu,\xi\nu,t)$ 
without explicitly knowing/extracting  the \mbox{$x$-dependence}  of the relevant GPD $H(x,\xi,t)$. 
Note that, due to the trigonometric weights, determining the CFF at low $\xi$ requires 
data at large $\nu$.

\begin{figure}
    \centering
    \includegraphics[width=0.95\linewidth]{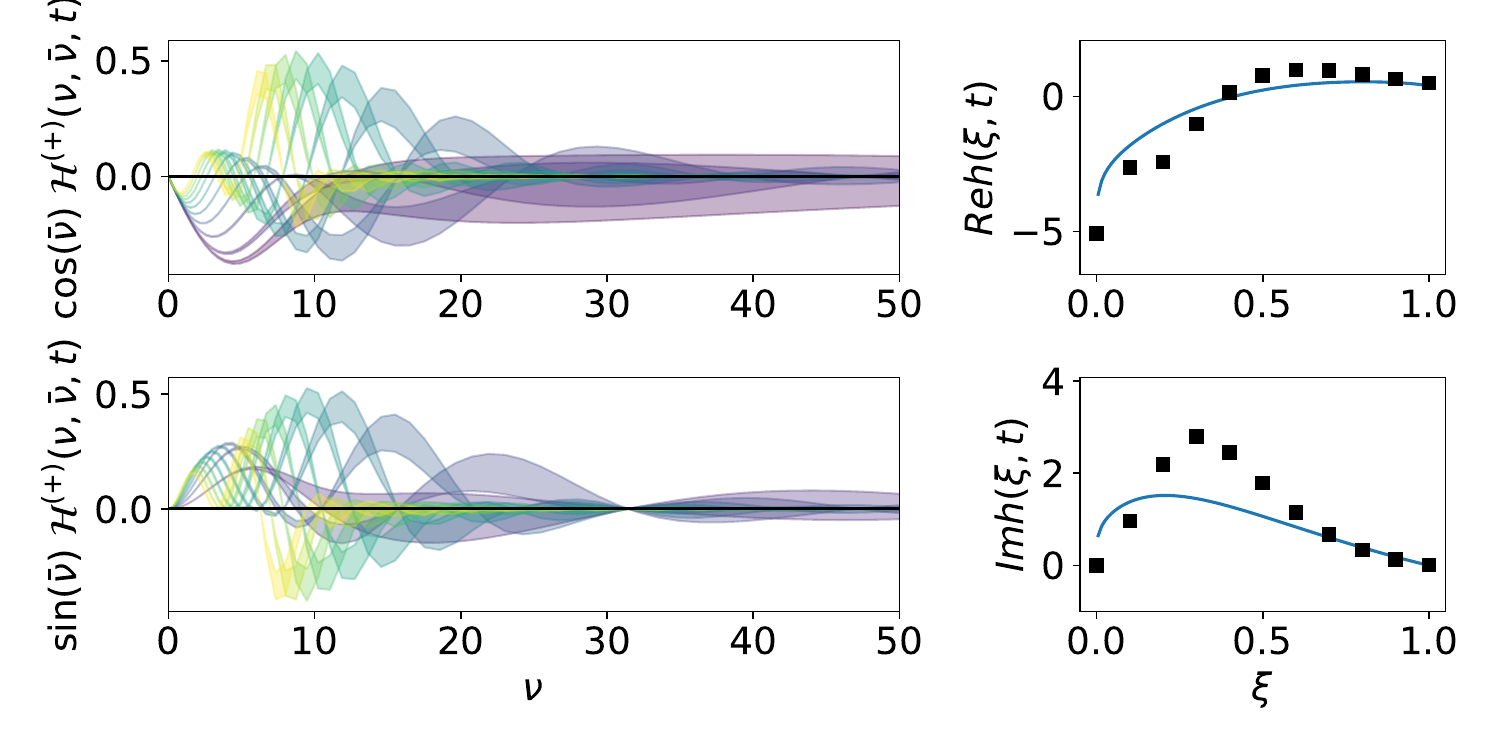}
    \caption{(Left) The integrand for the (Upper) real and (Lower) imaginary components of the iso-vector contribution to the CFF for $t=-0.8$ GeV${}^2$. The yellow band is $\xi=1$ then decreasing in steps of 0.1 until the violet $\xi=0$ band. (Right) The iso-vector contribution to the CFF from GK model (blue lines) and this analysis (black squares). }
    \label{fig:cff}
\end{figure}

In Fig.~\ref{fig:cff}, we show the integrands/integrals of the iso-vector contribution to the CFF in eqs.~\eqref{CFFpIm} and~\eqref{CFFpRe} calculated from the reanalyzed GPR. Reassuringly, the CFF determined from lattice data has the qualitative shape of the CFFs extracted from the GK model. 

Note that the Fourier factors amplify the relative weight  of the $\nu\in[5-20]$  region which, particularly at large $\xi$,  is where the GITD itself reaches a minimum. Future studies  with more systematic and statistical control over large momentum will be required for rigorous control over the extrapolation regime. 

The GPR model has substantial physics constraints by bounding the Fourier transform's domain and vanishing on the DD's boundary. Moreover, the DD model gives the correlation in $\nu$ and $\bar{\nu}$ that is encoded via polynomiality. Future studies can systematically test the  assumptions and choices of priors. 
In the Supplementary Materials, we illustrate the effect of different values of $t$.


\section{Conclusion}

 Hadrons are dynamical, strongly-coupled systems, and 
calculating their internal structure, specifically of protons, is a long-standing problem in particle and nuclear physics.  The study of  their three-dimensional structure in terms of quarks and gluons 
is a key goal for the experimental program of the 12~GeV  upgrade at Jefferson Lab and at the future EIC.   

This  structure is encoded in the  GPDs $H(x,\xi,t)$.  However,  extraction of GPDs
from experimentally  measurable   CFFs $  \mathfrak{H}(\xi,t)$
entails solving an ill-posed inverse problem.

Lattice QCD provides a unique opportunity  to calculate 
GPDs as  functions of all  three GPD variables.
This is achieved by  analyzing the  Ioffe-time distributions
$ \mathcal{H}(\nu,\xi\nu,t)$ that are directly 
 determined on the lattice. 
 These distributions capture the same dynamics as \mbox{$x$-dependent} distributions, but provide an intuitive and frame-independent space-time picture of hadrons.
 
In this work, we derived a direct expression for Compton Form Factors in terms of the Ioffe-time distribution 
 by-passing GPD extraction. This approach amounts to a new forward problem in the form of  a Fourier-like integral of the CP-odd GITD combination. We integrated a double distribution-based  model of the extracted Ioffe-time distribution  to obtain the first measurement of the CFF from lattice QCD and compared it to the CFF from the phenomenological GK model. In addition, we showed that the task of learning mechanical properties  of hadrons  turns into a problem of modeling the derivative of  GITD at $\nu=0$.

  We have introduced the impact-parameter Ioffe-time distribution, which describes space-time images of protons in terms of the transverse location ${\bf b}_T$ of the parton within the hadron and the length of interaction time $\nu /m_p$ with the probe.

We have calculated the 
Ioffe-time-dependent mean-squared radius of the proton by taking the derivative of the GITD at vanishing momentum transfer. 
By modeling lattice QCD data 
from~\cite{Dutrieux:2026grg}, we extracted the radii and compared the results with the transverse radii in the GK model. Working directly in 
Ioffe-time space enables radii from reliable interpolations compared to momentum fraction space providing higher fidelity lattice calculations. 

Improved knowledge of the radii can be obtained in future analyses by incorporating well-motivated models, such as modeling the momentum-transfer dependence using dipole fits. Moreover, future lattice QCD calculations, with controlled systematic uncertainties, will paint a more precise, more accurate, and more intuitive picture of the internal structure of hadrons, complementing the wealth of data that will be generated at Jefferson Lab and the EIC.

\section{Acknowledgments}
The authors gratefully acknowledge Herv\'e Dutrieux and C\'edric Mezrag for their collaboration during the initial stages of this work and for their critical reading of the manuscript.

This material is based upon work supported by the U.S. Department of Energy, Office of Science, Office of Nuclear Physics under Contract No. 89243126CSC000213.
SZ acknowledges support in part from the Agence Nationale de la Recherche
(ANR) under Project No. ANR-23-CE31-0019. CJM~is supported in part by U.S.~DOE Grant \mbox{\#DE-SC0025908}. KO is supported in part by U.S.~DOE Grant \mbox{\#DE-FG02-04ER41302}. The work of DR was conducted in
part under the Laboratory Directed Research and Development Program at
Thomas Jefferson National Accelerator Facility for the U.S.\ Department of Energy. AR acknowledges support by U.S.~DOE Grant \mbox{\#DE-FG02-97ER41028}. 
This project received funding from the European Research Council (ERC) via the project ``HADaSTRaLS” grant agreement 101231497. Funded by the European Union. Views and opinions expressed are however those of the author(s) only and do not necessarily reflect those of the European Union or the European Research Council Executive Agency (ERCEA). Neither the European Union nor the ERCEA can be held responsible for them.
 This work has benefited from the collaboration enabled by the Quark-Gluon Tomography (QGT) Topical Collaboration, U.S.~DOE Award \mbox{\#DE-SC0023646}. Computations for this work were carried out in part on facilities of the USQCD Collaboration, which are funded by the Office of Science of the U.S.~Department of Energy. The authors acknowledge William \& Mary Research Computing for providing computational resources and/or technical support, which were provided by contributions from the National Science Foundation (MRI grant PHY-1626177), and the Commonwealth of Virginia Equipment Trust Fund. In addition, this work used resources at NERSC, a DOE Office of Science User Facility supported by the Office of Science of the U.S. Department of Energy under Contract \#DE-AC02-05CH11231, as well as resources of the Oak Ridge Leadership Computing Facility at the Oak Ridge National Laboratory INCITE program, which is supported by the Office of Science of the U.S. Department of Energy under Contract No. \mbox{\#DE-AC05-00OR22725}. 
The authors acknowledge support as well as computing and storage resources by GENCI on Adastra (CINES), Jean-Zay (IDRIS) under project (2025-2026)-A0200516207.
The software codes
{\tt Chroma}~\cite{Edwards:2004sx}, {\tt QUDA}~\cite{Clark:2009wm,Babich:2010mu}, {\tt QUDA-MG}~\cite{Clark:SC2016}, {\tt QPhiX}~\cite{ISC13Phi},
{\tt MG\_PROTO}~\cite{MGProtoDownload}, {\tt QOPQDP}~\cite{Osborn:2010mb,Babich:2010qb}, and {\tt REDSTAR}~\cite{Chen:2023zyy,Selvitopi:2025mlx}
were used.
The authors acknowledge support from the U.S. Department of Energy, Office of Science, Office of Advanced Scientific Computing Research and Office of Nuclear Physics, Scientific Discovery through Advanced Computing (SciDAC) program, and of the U.S. Department of Energy Exascale Computing Project (ECP). The authors also acknowledge the Texas Advanced Computing Center (TACC) at The University of Texas at Austin for providing HPC resources, including the Frontera computing system~\cite{frontera} that has contributed to the research results reported within this paper.

\FloatBarrier
\bibliography{sample}

\appendix
\section{Supplementary Materials}
\section{$\nu$ and $\bar\nu$ dependence of the GITD}
The ITDs/GITDs $ \mathcal{H}^{(+)}(\nu,\xi\nu,t)$ can be formally written as a Taylor expansion in $\nu^n$, whose coefficients are proportional to the Mellin $x^n$-moments of the PDFs/GPDs~\cite{Ma:2017pxb,Karpie:2018zaz}. This observation  was exploited by ~\cite{HadStruc:2024rix,Gao:2025inf} to study the skewness dependence of the GPD's moments. One  can also write a double series   in $\nu,\bar{\nu}$
\begin{eqnarray}
    \mathcal{H}(\nu,\bar\nu) = \sum_n \frac{i^n}{n!} \bigg[ \sum_{k=0 }^{\lfloor{n}/2\rfloor} \bar\nu^{2k} \nu^{n-2k} A_{n+1,2k} \nonumber \\ +
    \frac{1-(-1)^n}{2}
    \frac{\bar\nu^{n+1}}{\nu} C_{n+1} \bigg]
\end{eqnarray}
where  $C_n$ is  the Mellin moments of the $D$ term of which 
the coefficients $A_{n,k}$ are  independent. Phenomenologically we are only interested in the region of $|\bar\nu|<|\nu|$ and not in a scenario with zero $\nu$ but  non-zero $\bar\nu$,  which would make the last  term divergent.

Similar to the PDF case, large $\nu$ asymptotics are instead decided by the behavior of $\beta$ as they approach their endpoint behavior. $\beta$ will diverge as it approaches 0, but remaining integrable means the rate is slower than $\beta^{-1}$. 
This will cause a power law behavior for large $\nu$ whose rate is slower than $\nu^{-1}$~\cite{Braun:1994jq}. Note that this guarantees that the integral in Eq.~\eqref{Kerrep2} convegent when $\epsilon\to0$. $\alpha$ on the other hand is convergent, meaning the large $\bar\nu$ dependence will decay more rapidly than $\nu$. This greater convergence is seen at fixed $\nu$ in the GPR reconstruction of Ref.~\cite{Dutrieux:2026grg}.

\section{More CFF plots}
\begin{figure}
    \centering
    \includegraphics[width=\linewidth]{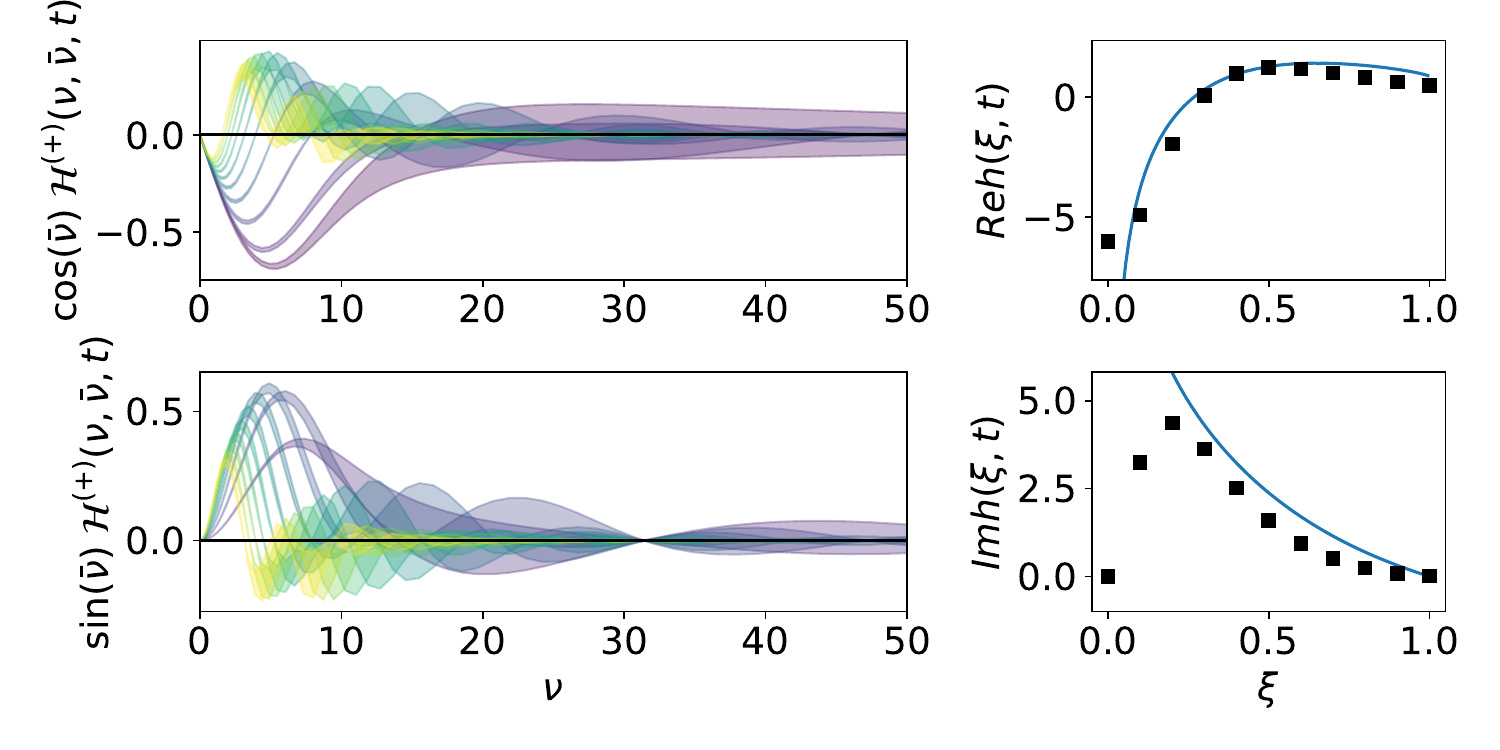}
    \caption{Same as Fig.~\ref{fig:cff} except at $t=0$ GeV${}^2$.}
    \label{fig:cff_extra1}
\end{figure}

\begin{figure}
    \centering
    \includegraphics[width=\linewidth]{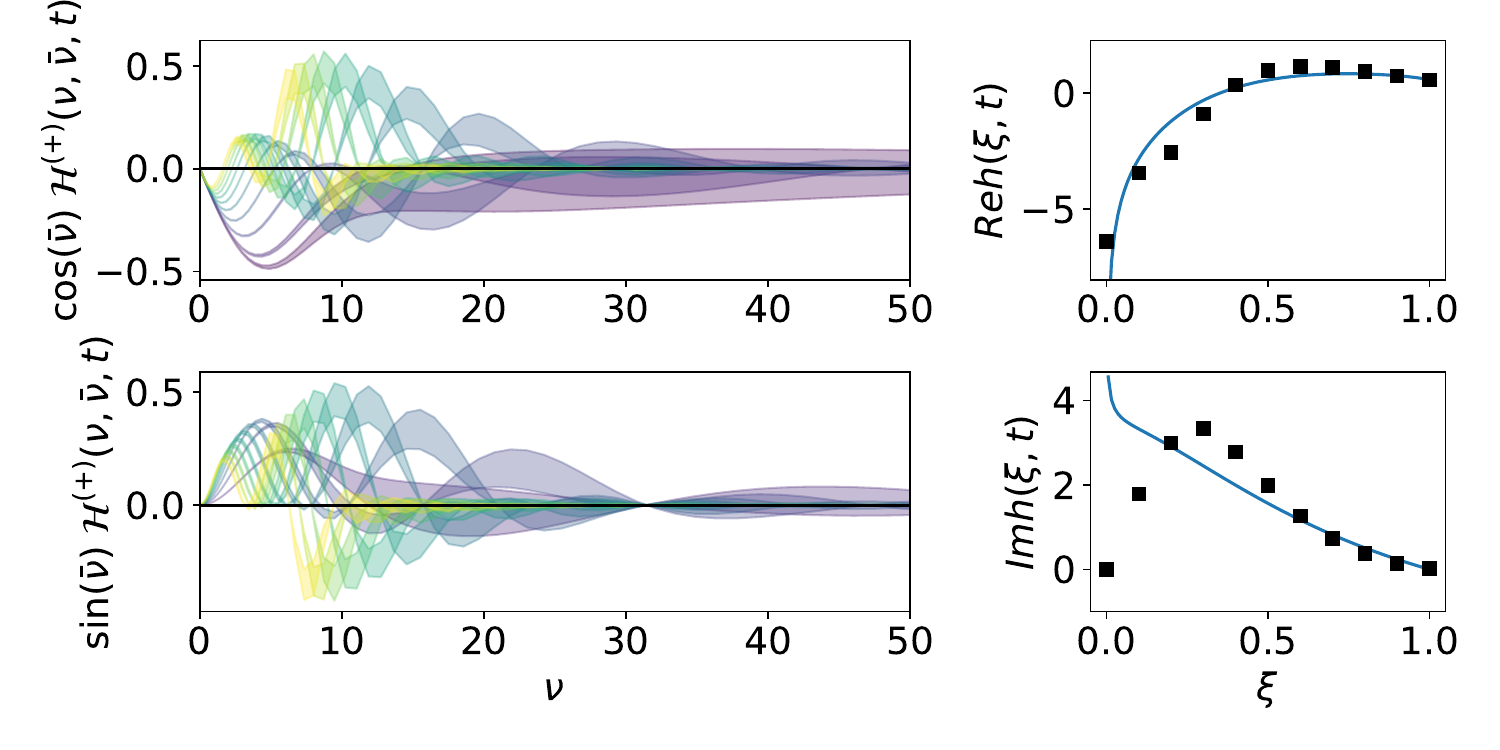}
    \caption{Same as Fig.~\ref{fig:cff} except at $t=-0.4$ GeV${}^2$.}
    \label{fig:cff_extra2}
\end{figure}

\begin{figure}
    \centering
    \includegraphics[width=\linewidth]{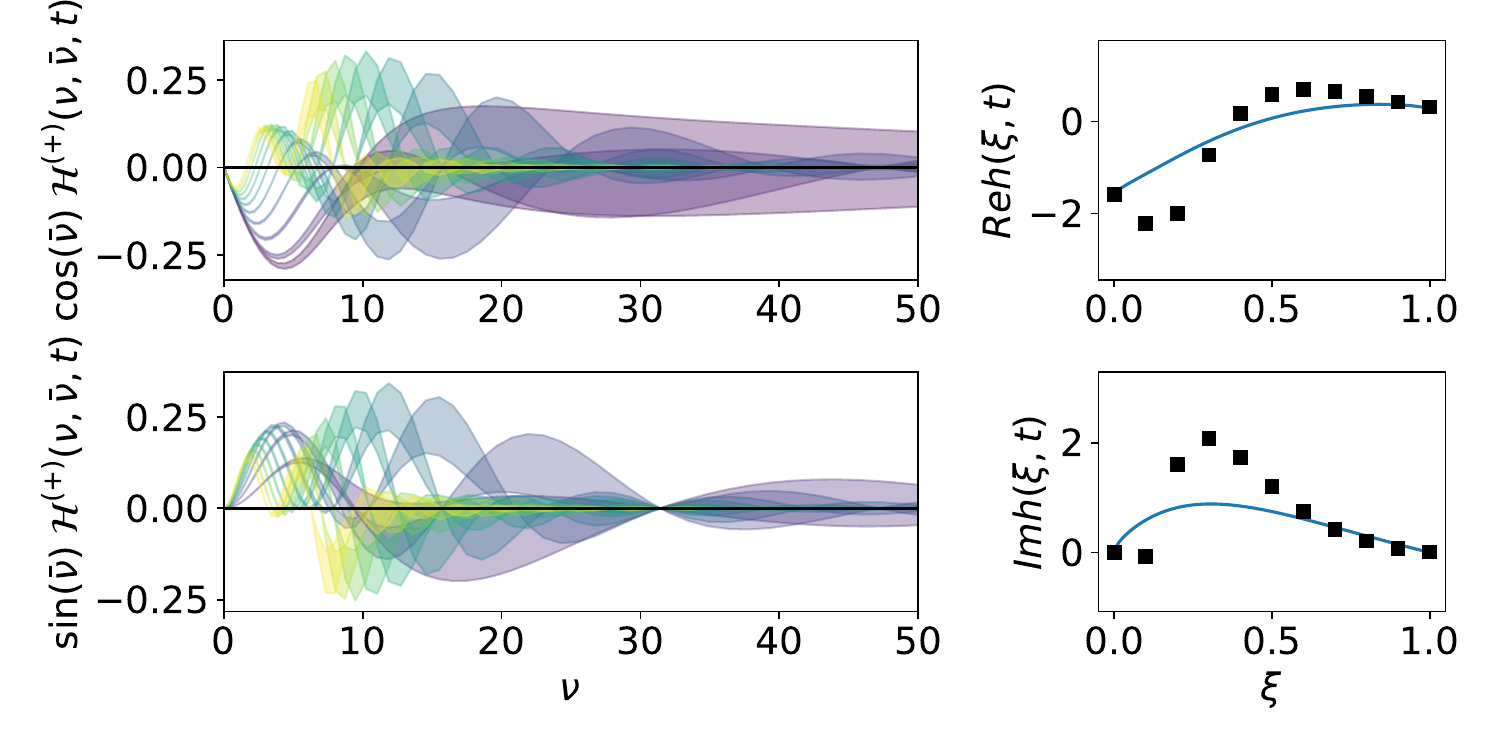}
    \caption{Same as Fig.~\ref{fig:cff} except at $t=-1.2$ GeV${}^2$.}
    \label{fig:cff_extra3}
\end{figure}

Figs.~\ref{fig:cff_extra1}-~\ref{fig:cff_extra3} show the integrand and integrals of Eqs~\eqref{CFFpIm} and~\eqref{CFFpRe}. For $t>-0.538$ GeV${}^2$, the GK model experiences a low $\xi$ divergence of the imaginary component. In this lattice calculation the low $t$ GPDs and therefore CFFs do not diverge in the imaginary component at small $\xi$ as the GK model does. It has been noticed in pseudo-PDF calculations~\cite{Orginos:2017kos} that at unphysical parameters the PDFs tend to not be divergent like the light cone PDF. Beyond simple model bias, one explanation which will be tested in future studies is the suppression of Regge behavior due to the unphysically heavy pion mass~\cite{Radyushkin:2018cvn,Radyushkin:2019mye}. 

\newpage

\section {Kullback-Leibler Divergence}

\begin{figure}[h]
    \centering
    \includegraphics[width=0.475\linewidth]{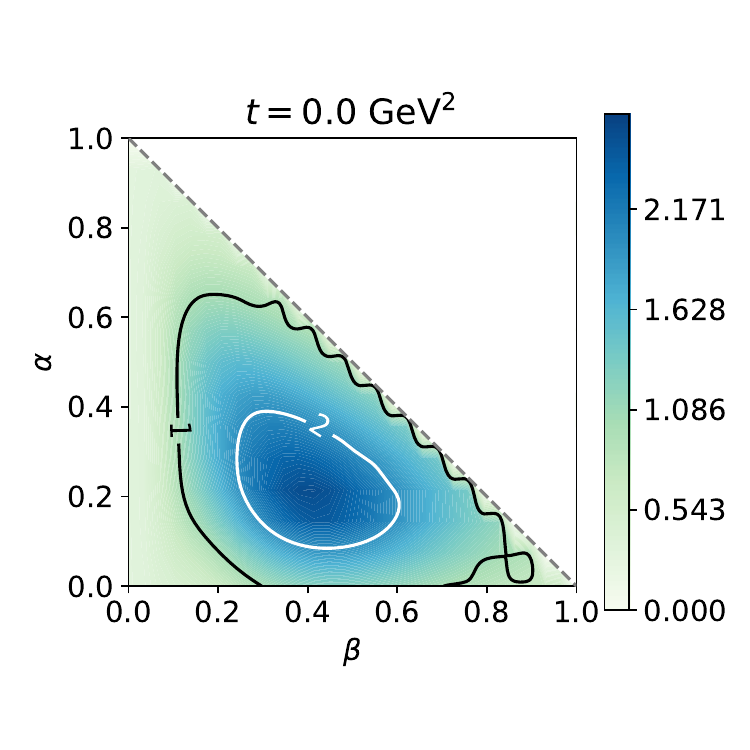}
    \includegraphics[width=0.475\linewidth]{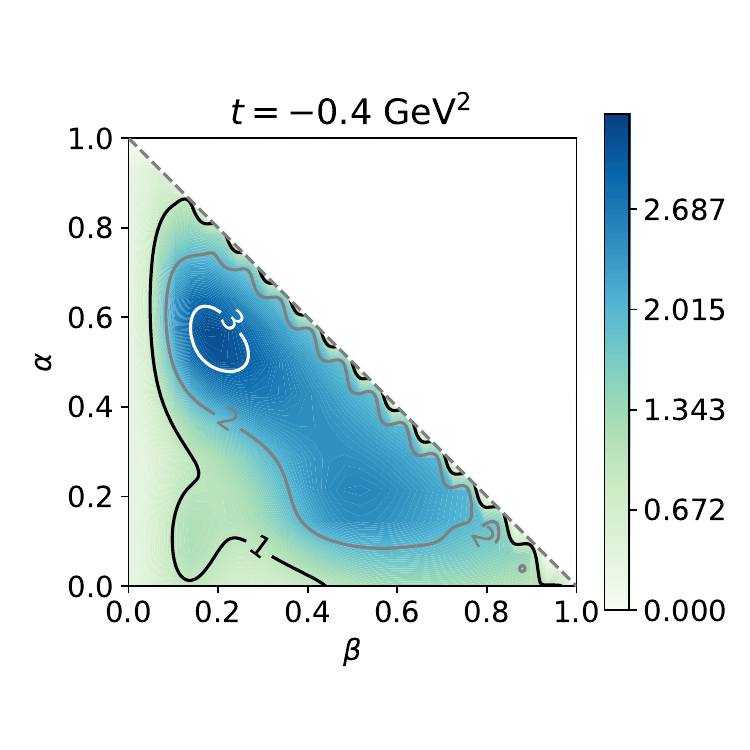}
    \includegraphics[width=0.475\linewidth]{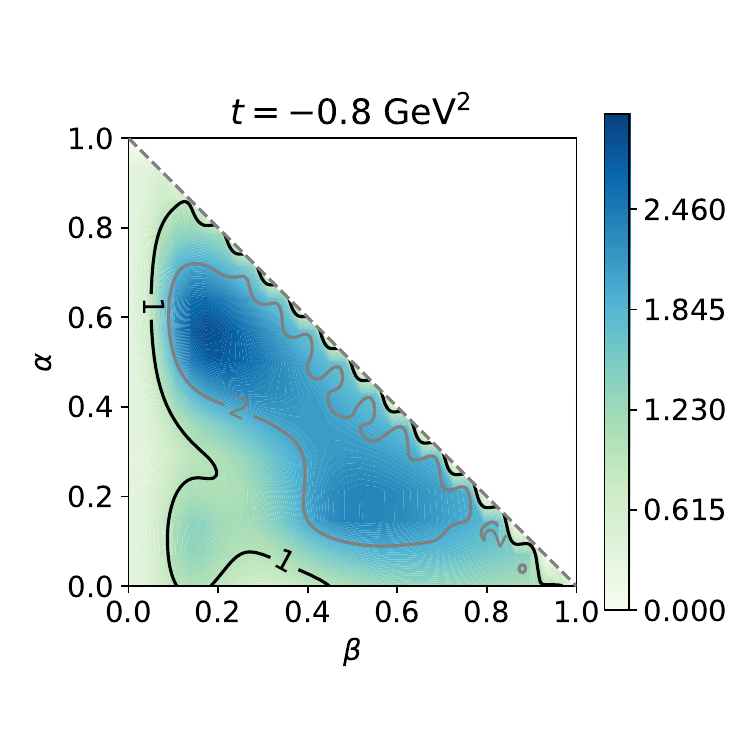}
    \includegraphics[width=0.475\linewidth]{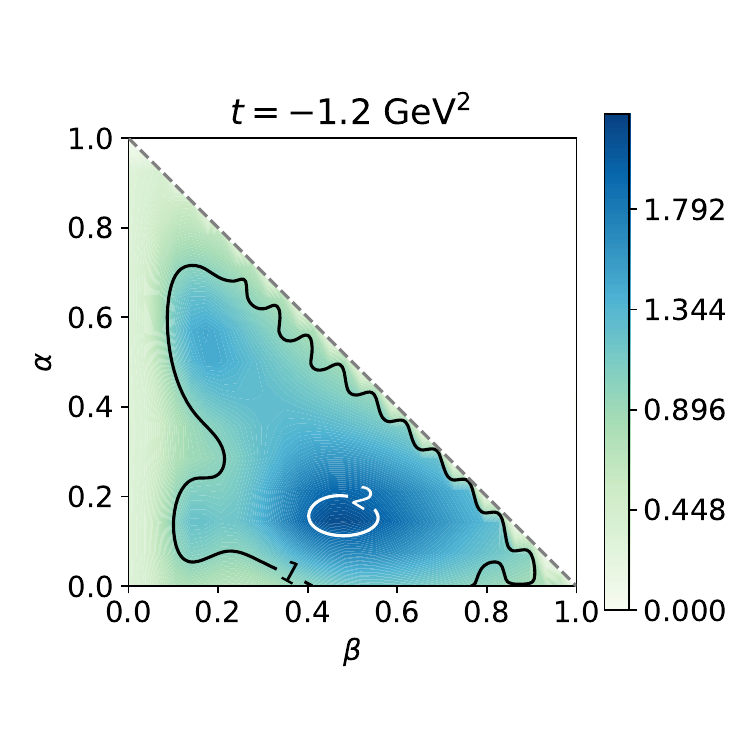}
    \caption{The KL divergence between the posterior and prior of the GPR for various $t$. Blue regions indicate places where the data had most impact on the posterior, i.e. where data made it differ the most from the prior. The total $D_{KL}$ of the analysis is 42.3.}
    \label{fig:dkl}
\end{figure}

The Kullback-Leibler (KL) Divergence~\cite{Kullback:1951zyt}, $D_{KL}$ is a measure of the differences between two distributions. The KL divergence can be calculated between the posterior and the prior distributions to estimate regions of the domain where the data impacted the result~\cite{ValentineSambridgeGP,Medrano:2025cmg}. The $D_{KL}$ for the reanalysis are shown in Fig.~\ref{fig:dkl}. 

This combined region of sensitivity from the different angled lines shows the way in which the DD space will correlate parts of the model, for example when extrapolating $\xi=1$ beyond the $\nu$ in the data. This $(\nu, \bar{\nu})$ extrapolation is done in a way completely consistent with the polynomiality even away from the data. In other words those extrapolated points are being informed by data at different $\xi$ with longer $\nu$ extent due to the polynomiality relationship. Still future study of more models with data that has more systematic control over the $(\nu,\bar{\nu})/(\nu,\xi)$ plane is necessary for robust phenomenology with lattice QCD models.

\end{document}